\documentclass{main}

\def\NIMA{{\em Nucl. Instrum. Methods} A}

\def\be{\begin{equation}}
\def\ee{\end{equation}}
\def\bea{\begin{eqnarray}}
\def\eea{\end{eqnarray}}

\usepackage{amsmath}
\usepackage{amssymb}

\newcommand{\Photo}{}

\begin{document}

\title{\Large STAR UPC Results}

\author{David Tlusty (for the STAR Collaboration)}

\address{College of Arts and Sciences, Creighton University.\\ Omaha, Nebraska, 68118 USA
}

\maketitle\abstracts{
Relativistic heavy ions are sources of strong electromagnetic fields which produce photon-induced interactions. 
These interactions are usually studied in ultra-peripheral collisions (UPCs) of the relativistic heavy ions. 
The UPCs can produce di-lepton or di-hadron pairs via the $\gamma-\gamma$ interactions or produce vector mesons 
via the $\gamma-$nuclear interactions. Both the photo-produced vector mesons and the lepton/hadron pairs productions 
depend on the original electromagnetic field. In addition, the photo-produced vector mesons are sensitive to the gluon 
parton distribution of the entire target nucleus (coherent production) or the individual nucleons (incoherent 
production). In these proceedings, recent STAR results on vector mesons, di-lepton pairs, and di-hadron 
photo-production will be discussed. The measurements are compared with available models to discuss the 
relevant implications. 
}


\section{Introduction}

Heavy ions, when accelerated to energies at which relativistic effects like length contraction or time dilation cannot be neglected,
generate very strong electromagnetic fields due to the Lorentz contraction and high charge. These fields are strong
enough to induce photo-nuclear and photon-photon interactions. The interaction cross sections are high enough, due to
large photon fluxes so that various experiments can be designed to study nuclear effect properties. Atomic nuclei consist 
of mutually bound protons and neutrons whose structure may break down further to valence quarks plus virtual gluons
and sea quarks and antiquarks during a heavy-ion collisions. The presence of sea quarks, antiquarks, and gluons fluctuates and depends on the kinematic
properties of a particular experiment. The deep inelastic scattering (DIS) experiments established the quark and gluons
presence through the parton (quark and gluon) distribution functions (PDFs) $q(x,Q^2)$ and $g(x,Q^2)$, where $x$ is the Bjorken momentum fraction - the fraction of the nucleon momentum carried by that parton, measured in the infinite momentum frame - and $Q^2$ is the momentum scale at which the measurement is made. 

The PDFs of nucleons bound in a nucleus are further affected
by the presence of other nucleons in close proximity. The reference \cite{RWang2017} shows a global analysis of many
DIS data with references to many relevant experiments collected over the years. At small Bjorken $x$ values 
($\lesssim 10^{-2}$) the parton densities are reduced; this process is called \emph{Nuclear Shadowing} and the 
effect is stronger with heavier nuclei. Around $x = 10^{-1}$, there is enhancement called 
\emph{Antishadowing}, and close to $x=1$, the EMC Collaboration discovered suppression \cite{emccol} and finally, 
at $x\sim 1$ we have an enhancement due to Fermi motion of nucleons.          

Photoproduction is an excellent tool to study nuclear effects, because
the production cross section is proportional to the square of the gluon density - at least two gluons are exchanged
in the photoproduction - and photon fluxes increase with the Lorentz contraction (EM field). Large powerful accelerators
like the LHC, RHIC, or SPS can then provide a large volume of data for photoproduction based experiments. Furthermore, modern
studies have gone beyond one-dimensional parton distribution, exploring spatial dependence of shadowing \cite{SpencerNature}.

This article discusses photoproduction in ultra-periperal heavy ion collisions (UPCs) at the Relativistic Heavy Ion 
Collider (RHIC) \cite{RHIC}. UPCs occur at impact parameters larger than the sum of the ions radii to avoid having hadronic interactions overshadow the electromagnetic processes. The photon flux scales as the square of the nuclear charge and the maximum photon energy is about $2\gamma^2\hbar c/R_A$, where $\gamma$ is the lab-frame beam Lorentz boost and $R_A$ is the 
radius of a nucleus. Most data collected at RHIC are from Au+Au collisions at $\sqrt{s}=200$ GeV which would generate the
maximum photon energy around 10 TeV. We recognize \emph{coherent} photoproduction where a projectile photon 
interacts with whole nucleus and the final state is exclusive; for example, the coherent photoproduction
of di-electron produces just an $e^+e^-$ pair and nothing else. Experimentally, the final state has very low $p_T$ which allows
easy separation from incoherent and hadronic interactions and the combinatorial background using kinematic cuts. The \emph{incoherent} 
photoproduction is a description of interactions where a photon interacts with particular nucleons. There could be
multiple interactions during one collision. Incoherent photoproduction is sensitive
to event-by-event fluctuations, whereas the coherent photoproduction is sensitive to the average gluon distributions.

The following sections describe the STAR experiment and the most recent experimental results.                    

\section{The STAR Experiment}

\begin{figure}[h]
\centering
\includegraphics[width=0.9\textwidth]{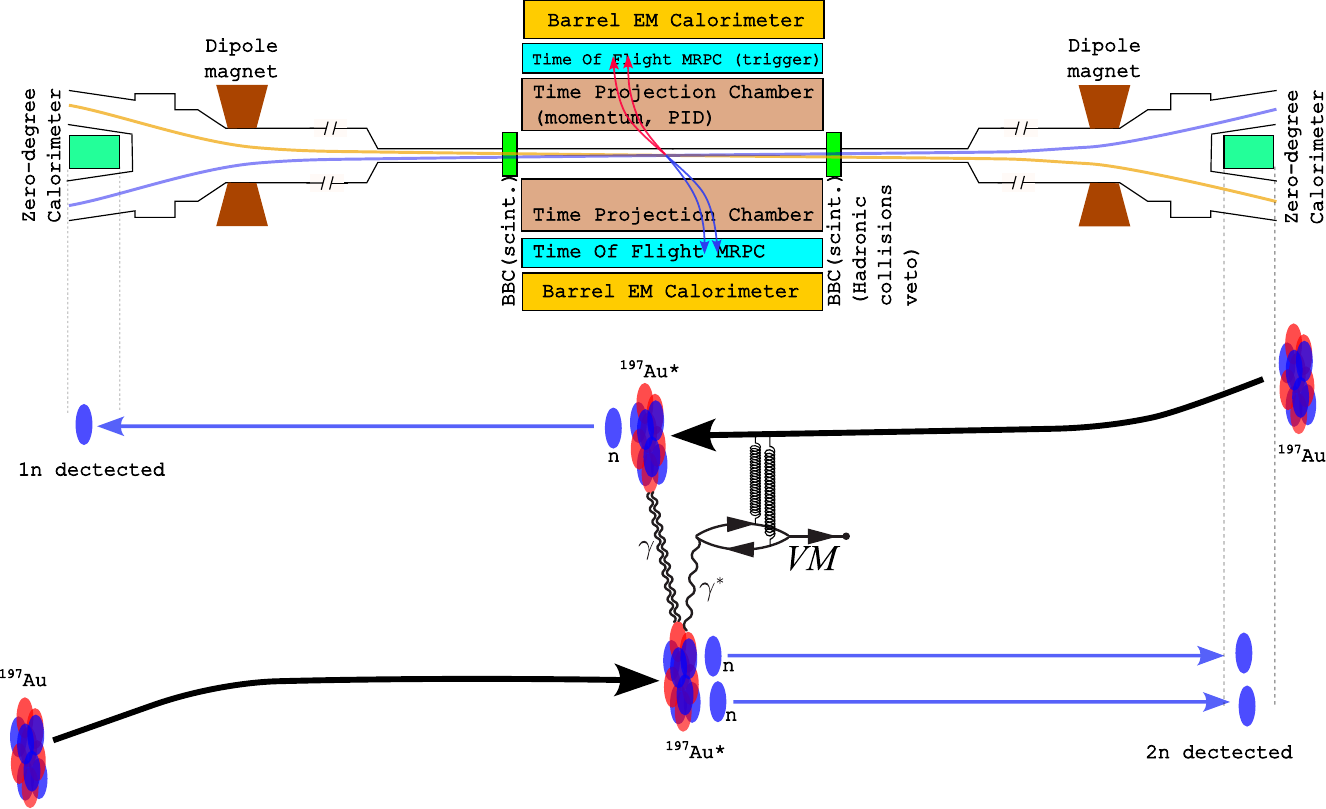}
\caption{Upper part: Simplified longituditnal view of the STAR experiment showing only the detector subsystems used in UPCs. 
Lower part: A scheme of vector meson production in UPC decaying into 4 (long living) daughter particles accompanied with 
mutual coulomb dissociation of gold nuclei.}     
\label{fig:STAR}
\end{figure}

The Solenoidal Tracker at RHIC (STAR) \cite{STARurl} is large acceptance detector with charged particle
tracking ability from rapidity $-1.5$ to $+1.5$ ($-1,1$ before 2019), electron calorimetry and muon detection 
from rapidity $-1$ to $+1$, 4 pairs of detectors in forward-backward rapidity, and newly installed forward tracking and calorimeter 
system. The UPC detection typically requires mid-rapidity tracking, neutron detection in zero-rapidity and 
detection veto in forward and backward rapidity region (see Figure \ref{fig:STAR}).       

\section{Latest Experimental Results}

\subsection{Vector Meson Production and Nuclear Imaging}\label{subsec:rho0}

The photoproduction of vector mesons can be modeled by the photon fluctuating to a quark-antiquark pair (see the lower part
of Figure \ref{fig:STAR}) which then scatters off the target nucleus (coherent) or a nucleon (incoherent), emerging as a real 
vector meson. For the coherent photoproduction in symmetric collisions, it is impossible to distinguish the target from the projectile.
This leads to a twofold ambiguity in photon energy and destructive interference between the two amplitudes at the lowest transverse
momentum of the final state. The nuclear structure is probed here at a scale set by the mass of the vector meson $M_V$:
$$Q^2 = (M_Vc^2/2)^2,$$
$$x = (M_Vc^2)^2/W^2,$$
where $W$ is the $\gamma$-nucleon centre-of-mass energy. 

\begin{figure}[h]
\centering
\includegraphics[width=0.45\textwidth]{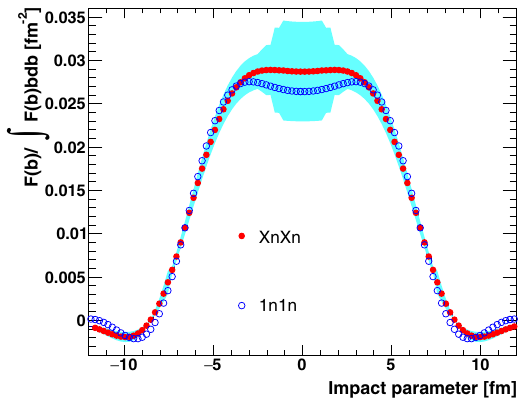}
\includegraphics[width=0.5\textwidth]{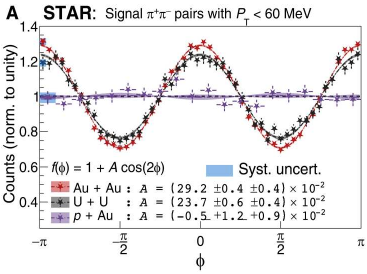}
\caption{Left: The target (Au nucleus) interaction elements distribution in the transverse plane, the result of a 2D Hankel 
transform of the $d\sigma_{xnxn}/dt$ and $d\sigma_{1n1n}/dt$ of $\rho_0$ production. The upper limit of the integral in equation 
(\ref{eq:Hankel}) is 0.06 GeV/$c^2$. Right: The $\phi$ distribution for $\pi^+\pi^-$ pairs produced in Au+Au and U+U collisions, 
the pair $p_T<0.06$ GeV and rest mass $0.65<M<0.9$ GeV.}     
\label{fig:NuclProfile}
\end{figure}

STAR paper \cite{rho0} describes in great detail the photoproduction
of $\rho_0$ mesons in Au+Au collisions at $\sqrt{s} = 200$ GeV. 1100 $\mu b^{-1}$ of UPC events were collected in 2010 
using a trigger based on nuclear dissociation (at least 1 neutron detected in both zero-degree calorimeters), a rapidity gap
veto to suppress hadronic contributions, and a final state multiplicity cut. $\rho_0$ were reconstructed from the hadronic decay 
channel ($\rho_0 \longrightarrow \pi^+\pi^-$). A nucleus can be imaged (in the transverse plane) by measuring the production 
cross section as a function of momentum transfer $d\sigma/dt$, where $t$ can be approximated by $p_T^2$. The Fourier transform 
of $d\sigma/dt$ gives the distribution of interaction elements $db$ of the nucleus. Gold nuclei are azimuthally symmetric, so
the radial distribution can be determined with a Hankel transformation:
\be
F(b) \equiv \frac{1}{2\pi}\int_0^\infty dp_Tp_TJ_0(bp_T)\sqrt{\frac{d\sigma}{dt}}.
\label{eq:Hankel}
\ee 
The results of the normalized profile is pictured in Figure \ref{fig:NuclProfile} Left. The full-width half-maximum (FWHM) of the
distribution is reported $2(6.17\pm0.12)$ fm in section \ref{subsec:rho0}. The tails are negative around $|b|=9$ fm which is due to interference between
the two nuclei (the paper \cite{rho0} only suggested this claim as a possibility, the confirmation was made later, discussed
in this section later).     

The previously mentioned interference has been discussed in the recently published paper \cite{rho0intf} as a result of an overlap of two
wave functions at a distance an order of magnitude larger than the $\rho_0$ travel distance within its lifetime. The produced vector 
mesons retain the incident photon quantum numbers, $J^{PC}=1^{--}$. When the spin-1 $\rho_0$ decays, its spin is transferred into 
the orbital angular momentum of the spin-0 pions, resulting in their momenta being preferentially aligned with the $\rho_0$ spin
direction. The angle $\phi$, defined as 
\be
\cos\phi=\frac{(\vec{p_{\pi^-}}+\vec{p_{\pi^+}})\cdot(\vec{p_{\pi^-}}-\vec{p_{\pi^+}})}{|\vec{p_{\pi^-}}+\vec{p_{\pi^+}}|\times|\vec{p_{\pi^-}}-\vec{p_{\pi^+}}|},
\label{eq:cosphi}
\ee      
was used as a proxy for the $\rho_0$ polarization direction relative to its transverse momentum. Figure \ref{fig:NuclProfile} shows a
clear $\cos(2\phi)$ modulation for Au+Au and U+U data, while p+Au data are consistent with an isotropic distribution of the daughter particles.

\begin{figure}[h]
\centering
\includegraphics[width=0.5\textwidth]{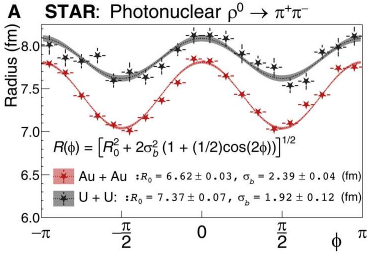}
\caption{Radial parameter as a function of the $\phi$ angle for Au+Au and U+U.}     
\label{fig:Radius}
\end{figure}

The data were split into $\phi$ bins their $t\equiv p_T^2$ distribution were fitted with function:

\be
f(t)=A_c\left| F[\rho_A(r;R,a)](|t|)\right|^2+\frac{A_i/Q_0^2}{(1+|t|/Q_0^2)^2},
\label{eq:FFFaktor} 
\ee 
where $A_c$ is the amount of coherent, and $A_i$ the amount of incoherent production. The $Q_0^2$ was fixed to $0.099\text{ GeV}^{-2}$ \cite{rho0intf}.
The $\rho_A(r;R,a)$ is the Woods-Saxon distribution 
$$ \rho_A(r;R,a) = \frac{\rho_0}{1+\exp[(r-R)/a]},$$
where $R$ is the nuclear radius, $a$ the surface thickness, and $\rho_0 = 3A/(4\pi R^3)$ geometric radius of the nucleus (Gold or Uranium). Figure
\ref{fig:Radius} shows the $R$ as a function of $\phi$ for Gold (Au) and Uranium (U) nuclei. To extract nucleus' true radius $R_0$, the interference
effect responsible for the modulation had to be subtracted, thus the equation:
$$R=\sqrt{R_0^2+\sigma_b^2/\varepsilon_p\times(1+\varepsilon_p\cos 2\phi)}$$
was used to fit the data in Figure \ref{fig:Radius}. The parameter $\varepsilon_p = \langle\cos 2\phi\rangle = 1/2$ accounts for the resolution
of the measurement when using the daughter momentum as a proxy for the polarization direction. The $\sigma_b$ quantifies the strength of the 
interference effect. The $R_0$ was found to be $6.62\pm0.03$ fm for the Au and $7.37\pm0.07$ fm for the U nucleus. The
Au nucleus measured by this interference method is consistent with the older measurement described in the previous section.    
   
\subsection{$J/\psi$ production and Nuclear Shadowing}\label{subsec:jpsi}
   
\begin{figure}[h]
\centering
\includegraphics[width=0.5\textwidth]{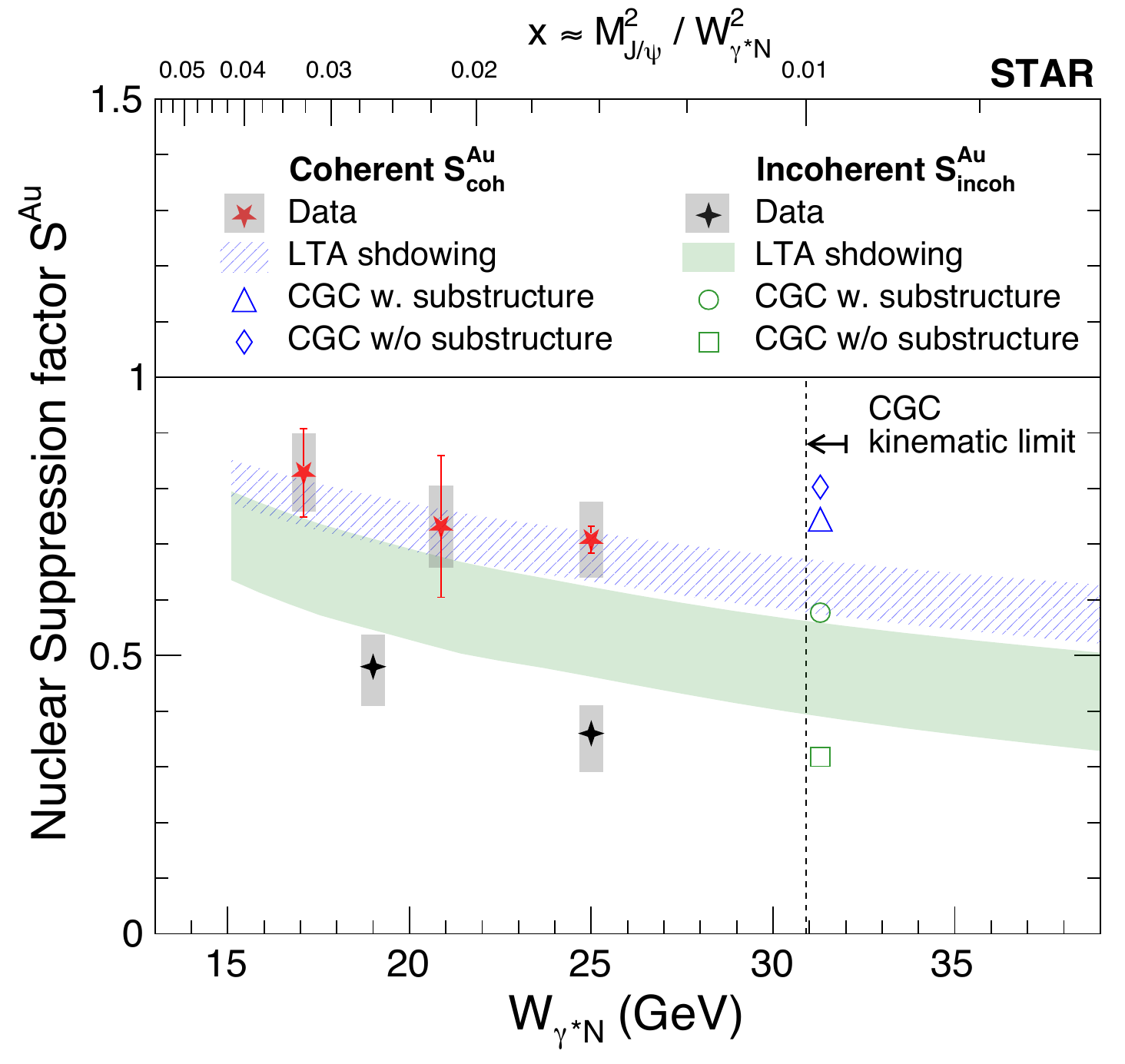}
\caption{Nuclear suppression factor of coherent and incoherent $J/\psi$ photoproduction in Au+Au UPCs. The data are compared with the nuclear shadowing model and the CGC model. The CGC poins are shifted from the vertical line for better visibility.}     
\label{fig:NuclearShadowing}
\end{figure}

Recently submitted papers \cite{jpsiSupression} and \cite{jpsiSupression2} present the exclusive $J/\psi, \psi(2S)$, and $e^+e^-$ pair photoproduction 
in Au+Au UPCs at $\sqrt{s_{NN}}=200$ GeV. The trigger setup allowed to distinguish three different neutron configurations, $0n0n, 0nXn$, and
$XnXn$. This separation allows us to resolve the photon energy ambiguity. The total cross section of coherent $J/\psi$ photoproduction 
in $\gamma$+Au collisions can be shown as a function of the photon-nucleon center-of-mass energy $W_{\gamma N}$. By comparing this cross section
with the Impulse Approximation (IA) \cite{ImpulseApproximation}, the nuclear suppression factor of coherent $J/\psi$ photoproduction, $S^\text{Au}_\text{coh}$, 
was calculated. Moreover, incoherent $J/\psi$ photoproduction was measured up to $p_T^2 = 2.2$ GeV/$c^2$ and the nuclear suppression factor if incoherent
$J/\psi$ photoproduction, $S^\text{Au}_\text{incoh}$, was calculated by comparing the incoherent production cross section with the free proton data at HERA
\cite{HERA}. Both $S^\text{Au}_\text{coh}$ and $S^\text{Au}_\text{incoh}$ are shown in Figure \ref{fig:NuclearShadowing}, together with the Leading Twist Approximation
(LTA) \cite{NuclShadowingModel} and the Color Glass Condensate (CGC) \cite{CGCmodel} models. The shadowing of the coherent production shows consistency with
the LTA while the incoherent production is suppressed more.

The papers \cite{jpsiSupression} and \cite{jpsiSupression2} further present the first $\psi(2S)$ photo-production measurement at RHIC.
Both coherent $J/\psi$ and $\psi(2S)$ photo-production cross sections were compared with the Next-to-Leading Order pQCD prediction
based on the nuclear PDF EPPS21 \cite{NLO}. The description of the data is off by a factor two at midrapidity, which may indicate
the large uncertainty on the nuclear parton distribution functions that this data can significantly constrain. The nuclear parton 
density at the top RHIC energy is in the region between large momentum quarks ($x_\text{parton}>0.1$) and low momentum gluons  
($x_\text{parton}<0.001$), which is essential to the understanding of nuclear modification effects in this transition regime. 

The quantum interference effect in symmetric Au+Au collisions was studied 
in $J/\psi$ photoproduction. The interference effect was observed in mid rapidity bins ($0.0<|y|<0.2$, $0.2<|y|<0.5$),
but not outside mid rapidity ($0.5<|y|<1.0$).  

The QED process $\gamma\gamma\rightarrow e^+e^-$ has been measured up to an invariant mass of 6 GeV/$c^2$ for different neutron emission classes,
which constrains the modeling of neutron emission and photon flux. The data provide important constrains to the parton density and its 
fluctuations, and also provide an essential experimental baseline for such measurement at the upcoming Election-Ion Collider. 

\subsection{First Observation of Breit-Wheeler Process}\label{subsec:breitw}

The paper \cite{BreitW} published in 2021 presents a measurement of $e^+e^-$ momentum and angular distributions from linearly polarized 
photon collisions. In QED, $\gamma\gamma\rightarrow e^+e^-$ processes are classified depending on the virtuality of the photons and
on whether the consideration of higher-order processes is necessary. Besides the collision of two virtual photons (Landau process) and one virtual and
one real photon (Bethe-Heitler process), there is the collision of two real photons, called the Breit-Wheeler process \cite{BreitWprocess}. 

The analysis included the total $\gamma\gamma\rightarrow e^+e^-$ production rate, the photon energy spectrum with sufficient precision 
with the initial spatial distribution of the electromagnetic field, and the allowed helicity states for participating photons via
measurement of the invariant mass spectra to demonstrate the absence of vector mesons. Furthermore, the analysis shows the measurement 
of the unique  $\cos(4\Delta\phi)$ modulation predicted for the Breit-Wheeler photon-photon fusion process to demonstrate
that the interacting photons are real and transversely linearly polarized. 

The most common background process is the photonuclear production of the $\rho_0$ decaying to $\pi^+\pi^-$. For this reason, high purity 
identification of $e^+e^-$ pairs is crucial. By combination of ionization loss and time-of-flight measurement, better than 99\% pure
$e^+e^-$ selection was achieved.  

\begin{figure}[h]
\centering
\includegraphics[width=\textwidth]{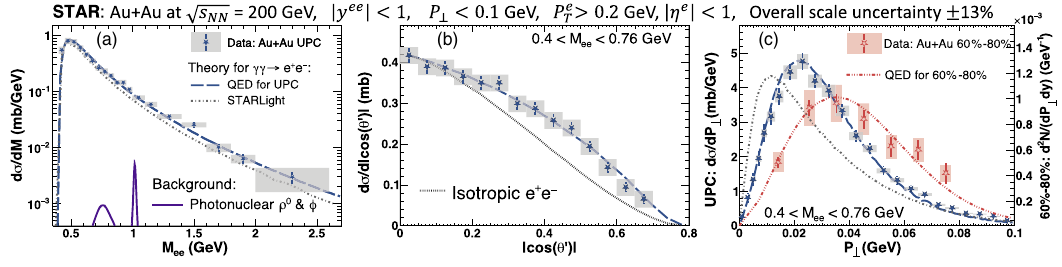}
\caption{The fully corrected differential cross sections for exclusively produced $e^+e^-$ pairs with respect to (a) the invariant mass $M_{e^+e^-}$, (b) the polar angle distribution $|\cos\theta'|$, and (c) the pair transverse momentum $P_T$}     
\label{fig:BW1}
\end{figure}   

Figure \ref{fig:BW1}(a) shows the invariant mass of exclusive $e^+e^-$. The spectrum is smooth, absent of any vector mesons. This is a consequence
of the quantum numbers of the two photons involved in the Breit-Wheeler process where helicity state $J_z = 0$ is absent for real photons, 
but necessary for exclusive single vector-meson production. Figure \ref{fig:BW1}(b) shows the $|\cos\theta'|$ distribution, in which $\theta'$
is the polar angle of the $e^+$ momentum vector with respect to the beam, measured in the $e^+e^-$ center-of-mass frame. The gross detector
acceptance limits detection of particles to $45^\circ\lesssim\theta\lesssim 135^\circ$ which affects the $|\cos\theta'|$ shape. However, the
Breit-Wheeler process exhibits an enhancement toward a small polar angle to that of isotropic $e^+e^-$ prediction. Figure \ref{fig:BW1}(c)
shows differential cross section as a function of the pair transverse momentum $P_T$ in UPCs compared with the same distribution in 60\%-80\% central
collisions to demonstrate the sensitivity of the process to the initial geometry of the colliding electromagnetic fields. 

\begin{figure}[h]
\centering
\includegraphics[width=0.46\textwidth]{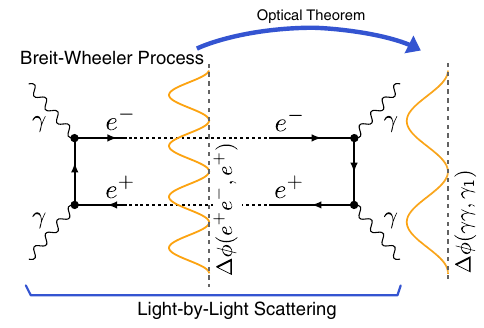}
\includegraphics[width=0.46\textwidth]{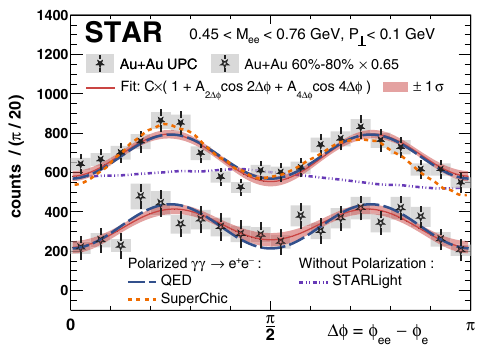}
\caption{Left: A Feynman diagram for the exclusive Breit-Wheeler process and the related light-by-light scattering process illustrating the unique angular distribution predicted for each process due to the initial photon polarization. Right: The $\Delta\phi = \phi_{ee}-\phi_e$ distribution from UPCs and 60\%-80\% central collisions for $M_{ee}>0.45$ GeV witch calculation from QED, STARLight, and SUPERCHIC3.}     
\label{fig:BW2}
\end{figure} 

According to the optical theorem, the Breit-Wheeler process and the $e^+e^-$ channel of light-by-light scattering, shown in the left panel of Figure \ref{fig:BW2}, are two parts of the same process. The right panel of Figure \ref{fig:BW2} Right shows the first measurement of the angular distribution $\Delta\phi$ for $e^+e^-$ pairs produced in $\gamma\gamma$
collisions, where $\Delta\phi$ is the azimuthal angle in the laboratory frame between the momentum of the $e^+e^-$ pair and one of the daughters.
Distribution from both UPCs and 60\%-80\% central collisions are shown with fits to a function of the form
\be
f(\Delta\phi)=C\left(1+A_{2\Delta\phi} \cos(2\Delta\phi)+A_{4\Delta\phi} \cos(4\Delta\phi)\right),
\label{eq:BWmod} 
\ee 
where $C$ is a constant and $A_{2\Delta\phi}(A_{4\Delta\phi})$ is the magnitude of a $\cos 2\Delta\phi (\cos 4\Delta\phi)$ modulation.  
In UPCs, the $\cos 4\Delta\phi$ modulation is observed with an amplitude of $(16.8\pm2.5)\%$, which is in good agreement with numerical 
lowest-order QED calculations predicting an amplitude of 16.5\%. The data are also compared to predictions from the STARLight and SUPERCHIC  models. 
STARLight does not employ any polarization-dependent effects, and predicts an isotropic distribution. SUPERCHIC is a model similar to STARLight, but
with the photon helicity dependence determined by the orientation of the electromagnetic fields in the transverse plane. 

\section{STAR Forward Upgrade}

\begin{figure}[h]
\centering
\includegraphics[width=\textwidth]{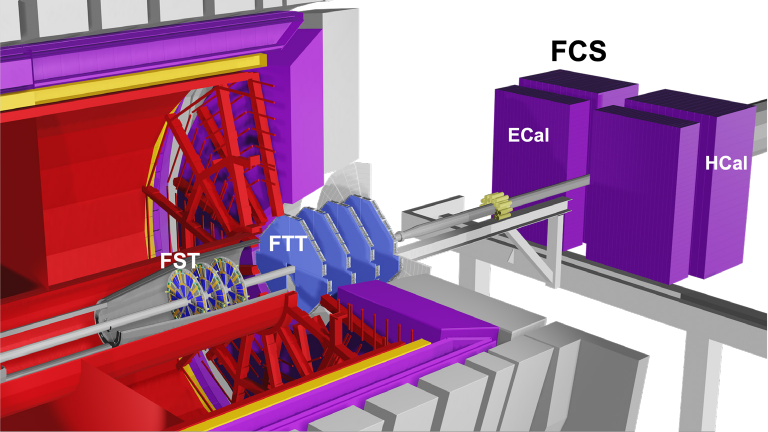}
\caption{The sketch of the STAR detector with the forward upgrade. This sketch was created using Blender and is based on the actual geometry of the STAR detector. The components of the STAR forward upgrade are labeled.}     
\label{fig:forward}
\end{figure}   

Since 2022, STAR has been equipped with forward detectors covering the pseudorapidity region of $2.5<\eta<4.0$.
They consist of:
\begin{itemize}
\item the Forward Silicon Tracker (FST) - 3 planes of silicon strip detector
\item the Forward sTGC Tracker (FTT) - 4 planes of gas wire chambers 
\item the Forward Electromagnetic Calorimeter (ECal)
\item the Forward Hadron Calorimeter (HCal)
\end{itemize} 

The CAD picture of the STAR Forward Upgrade is shown in Figure \ref{fig:forward}. The forward upgrade allows STAR to 
extend its Bjorken $x$ coverage. It will allow STAR to analyze $\phi$ meson photoproduction for the first time, spin-dependent
vector meson production and extent the photon-nucleon center-of-mass energy to below 10 GeV.

\section*{Acknowledgments}

This work was supported in part by the Office of Nuclear Physics within the U.S. DOE Office of Science.


\vspace{10cm}


\section*{References}

\begin{thebibliography}{99}

\bibitem{RWang2017}
R. Wang, X. Chen, and Q. Fu, 
Nucl. Phys. B \textbf{920}, 1 (2017)
doi:10.1016/j.nuclphysb.2017.04.008
\bibitem{emccol}
N. Burtebayev, 
Int. Jour. Mod. Phys. E \textbf{27}, 1850025 (2018) 
doi:10.1142/S0218301318500258
\bibitem{SpencerNature}
S. Klein and H. M\"antysaari, 
Nat. Rev. Phys. \textbf{1}, 662-674 (2019) 
doi:10.1038/s42254-019-0107-6
\bibitem{RHIC}
H. Hahn et al,
\NIMA \textbf{499}, 2-3, 245-263 (2003)
doi:10.1016/S0168-9002(02)01938-1
\bibitem{STARurl}
url:www.star.bnl.gov
\bibitem{rho0}
L. Adamczyk et al. (STAR Collaboration),
Phys. Rev. C \textbf{96}, 054904 (2017)
doi:10.1103/PhysRevC.96.054904
\bibitem{rho0intf}
L. Adamczyk et al. (STAR Collaboration),
Sci. Adv. \textbf{9}, abq3903 (2023)
doi:10.1126/sciadv.abq3903
\bibitem{jpsiSupression}
M. I. Abdulhamid et al. (STAR Collaboration),
arXiv:2311.13632 (2023)
\bibitem{jpsiSupression2}
M. I. Abdulhamid et al. (STAR Collaboration),
arXiv:2311.13637v2 (2023)
accepted by PRL, Jun 26th 2024 
\bibitem{NuclShadowingModel}
E. Kryshen, M. Strikman, and M. Zhalov,
arXiv:2303.12052 (2023)
\bibitem{CGCmodel}
H. M\"antysaari, F. Salazar, and B. Schenke,
arXiv:2207.03712 (2022)
\bibitem{ImpulseApproximation}
V. Guzey, E. Kryshen, M. Strikman, and M. Zhalov,
Phys. Lett. B \textbf{726}, 290 (2013)
doi:10.1016/j.physletb.2013.08.043
\bibitem{HERA}
C. Alexa et al. (H1 Collaboration),
Eur. Phys. J. C \textbf{73}, 2466 (2013)
doi:10.1140/epjc/s10052-013-2466-y
\bibitem{NLO}
K. J. Eskola, P. Paakkinen, H. Paukkunen, and C. A.Salgado,
SciPost Phys. Proc. \textbf{8}, 033 (2022)
doi:10.21468/SciPostPhysProc.8.033
\bibitem{BreitW}
J. Adam et al. (STAR Collaboration)
Phys. Rev. Lett. \textbf{127}, 052302 (2021)
doi:10.1103/PhysRevLett.127.052302
\bibitem{BreitWprocess}
G. Breit and J. A. Wheeler,
Phys. Rev. \textbf{46}, 1087 (1934)
\bibitem{BWQED}
W. Zha, J.D. Brandenburg, Z. Tang, and Z. Xu, 
Phys. Lett. B \textbf{800}, 135089 (2020)
doi:10.1016/j.physletb.2019.135089
\bibitem{starlight}
S.R. Klein, J. Nystrand, J. Seger, Y. Gorbunov, and J. Butterworth,
Comput. Phys. Commun. \textbf{212}, 258 (2017)
doi:10.1016/j.cpc.2016.10.016
\bibitem{superchic}
L. A. Harland-Lang, V. A. Khoze, and M. G. Ryskin,
Eur. Phys. J. C \textbf{79}, 39 (2019)
doi:10.1140/epjc/s10052-018-6530-5

\end{thebibliography}
\end{document}